\documentclass[12pt]{article}
\usepackage[centertags]{amsmath}
\usepackage{amssymb}
\usepackage{amsthm}
\usepackage{enumerate}

\usepackage{multirow}
\usepackage{graphicx}

\makeatletter
  \def\tagform@#1{\maketag@@@{(#1)\@@italiccorr}}
\makeatother

\begin{document}

\title{Generalized derivations and general relativity}

\author{Michael Heller \\
Copernicus Center for Interdisciplinary Studies, Cracow, Poland
\and Tomasz Miller\thanks{Corresponding author. E-mail: T.Miller@mini.pw.edu.pl} \quad \ Leszek Pysiak \quad \ Wies{\l}aw Sasin\\
Faculty of Mathematics and Information Science, \\ Warsaw
University of
Technology \\
ul. Koszykowa 75, 00-662 Warsaw, Poland \\
and Copernicus Center for Interdisciplinary Studies, Cracow, Poland}
\date{\today}
\maketitle

\begin{abstract}
We construct differential geometry (connection, curvature, etc.) based on generalized derivations of an~algebra ${\cal A}$. Such a~derivation, introduced by Bre\v{s}ar in 1991, is given by a~linear mapping $u: {\cal A} \rightarrow {\cal A}$ such that there exists a~usual derivation $d$ of ${\cal A}$ satisfying the~generalized Leibniz rule $u(a b) = u(a) b + a \, d(b)$ for all $a,b \in \cal A$. The generalized geometry ``is tested'' in the~case of the~algebra of smooth functions on a~manifold. We then apply this machinery to study generalized general relativity. We define the~Einstein--Hilbert action and deduce from it Einstein's field equations. We show that for a~special class of metrics containing, besides the~usual metric components, only one nonzero term, the~action reduces to the~O'Hanlon action that is the~Brans--Dicke action with potential and with the~parameter $\omega$ equal to zero. We also show that the~generalized Einstein equations (with zero energy--stress tensor) are equivalent to those of the~Kaluza--Klein theory satisfying a~``modified cylinder condition'' and having a~noncompact extra dimension. This opens a~possibility to consider Kaluza--Klein models with a~noncompact extra dimension that remains invisible for a~macroscopic observer. In our approach, this extra dimension is not an~additional physical space--time dimension but appears because of the~generalization of the~derivation concept.
\\

\end{abstract}

\maketitle

\section{Introduction}

In the~present paper we investigate differential geometry based on generalized derivations introduced in 1991 by Bre\v{s}ar \cite{Bresar}. He originally used them in his algebraic research concerning a~certain generalization of Posner's theorem \cite{Posner}. Systematic studies of algebraic properties of generalized derivations were initiated in 1998 by Hvala \cite{Hvala}. Since then, generalized derivations have been thoroughly studied by numerous re\-searchers \cite{Quadri,Ali,Nakajima1} and the~concept itself was further developed to encompass e.g. higher order derivations \cite{Nakajima2} and nonassociative settings \cite{Leger}. For a~brief summary and further references, see Ashraf \textit{et al.} \cite{OnDerivations}. However, as far as we know the~geometric content of this notion has not yet been investigated. The aim of the~present paper is to develop elements of differential geometry based on the~concept of generalized derivations and to see how this geometry works in the~context of general relativity theory.

After briefly presenting the~generalization itself (section \ref{sec2}), we construct basic notions of differential geometry based on this generalization (section \ref{sec3}), and apply them to the~case of algebra of smooth functions on a~lorentzian manifold (section \ref{sec4}). Because the~general case leads to rather involved calculations, we specify to the~case of a~simplified metric with only one additional nonzero term (section \ref{sec5}). We apply the~generalized geometry to formulate a~generalized theory of relativity (section \ref{sec6}). We start with a~natural choice of the~Einstein--Hilbert action and deduce from it the~generalized Einstein equations. They involve no free parameters. A term modeling the~space--time dependence on the~gravitational ``constant'' $G$ leads to similar effects as the~ones in Brans--Dicke theory \cite{Brans}. In fact, we show that for a~special class of metrics discussed in section \ref{sec5}, the~action reduces to the~O'Hanlon action \cite{OHanlon}, that is the~Brans--Dicke action with potential and with the~Brans--Dicke parameter $\omega$ equal to zero.

Even at first glance, the~generalized de\-ri\-va\-tion-based approach to general relativity seems to resemble the~idea standing behind Kaluza--Klein--type theories. We show (section \ref{sec7}) that indeed this observation is correct and the~generalized Einstein equations (with zero energy--stress tensor) can be equivalently obtained from a~Kaluza--Klein theory involving a~modified version of the~``cylinder condition''. However, unlike in standard gravity theories with extra dimensions, this equivalent Kaluza--Klein theory features a~noncompact extra dimension. The generalized general theory of relativity may thus serve as an~alternative formulation of a~Kaluza--Klein theory with a~single noncompact extra dimension that is not associated with any extra space--time dimension. This effectively avoids the~conundrum: why is this extra dimension not physically observed?

Let us finally mention that there exist many generalizations of standard geometry, of which the~most renowned is the~one developed by Connes and his collaborators (see for instance the~monographs \cite{Connes, Madore, Bondia}). The present work can be situated in the~stream of derivation-based approaches developed by Dubois-Violette \cite{Dubois-Violette1,Dubois-Violette2,Djemai}.

\section{Generalized derivations}
\label{sec2}

Throughout this section, $\cal A$ denotes an~(abstract) associative algebra over $\mathbb{K} = \mathbb{R}$ or $\mathbb{C}$. The algebra $\cal A$ can in general be nonunital and noncommutative. ${\cal Z}({\cal A})$ denotes the~center of the~algebra $\cal A$.

To begin with, let us recall that a~linear mapping $d: {\cal A} \rightarrow {\cal A}$ is called a~\textit{derivation} if it satisfies the~Leibniz rule: $d(a b) = d(a)b + a d(b)$ for all $a,b \in \cal A$. The set of all derivations of $\cal A$ is denoted $\textnormal{Der}({\cal A})$.

Derivations have the~following four properties, indispensable for the~deriv\-ation-based approach to differential geometry
\begin{enumerate}[(i)]
\item $\forall \, d_1, d_2 \in \textnormal{Der}({\cal A}) \ \forall \, \lambda_1, \lambda_2 \in \mathbb{K} \quad \lambda_1 d_1 + \lambda_2 d_2 \in \textnormal{Der}({\cal A})$,
\item $\forall \, d_1, d_2 \in \textnormal{Der}({\cal A}) \quad [d_1, d_2] \in \textnormal{Der}({\cal A})$,
\item $\forall \, d \in \textnormal{Der}({\cal A}) \ \forall \, f \in {\cal Z}({\cal A}) \quad f d \in \textnormal{Der}({\cal A})$,
\item $\forall \, d \in \textnormal{Der}({\cal A}) \ \forall \, f \in {\cal Z}({\cal A}) \quad d(f) \in {\cal Z}({\cal A})$.
\end{enumerate}

By (i, ii) $\textnormal{Der}({\cal A})$ possesses the~Lie algebra structure. By (i, iii), it is also a~${\cal Z}({\cal A})$-module. Finally, (iv) states that derivations leave the~center of $\cal A$ invariant.

By \textit{inner derivation} induced by $a \in {\cal A}$ we mean a~derivation $\textnormal{ad}_a(b) = [a,b] = ab - ba$ for any $b \in {\cal A}$. The set of all inner derivations is denoted $\textnormal{Inn}({\cal A})$.

In their 1991 paper \cite{Bresar}, Bre\v{s}ar considered what was called a~\textit{generalized inner derivation}, that is a~map $I_{a,b}: {\cal A} \rightarrow {\cal A}$ given by
\begin{align*}
\forall \, x \in {\cal A} \quad I_{a,b}(x) = ax + xb.
\end{align*}

Of course, $\textnormal{ad}_a = I_{a, -a}$. One can also easily notice that $I_{a,b}$ satisfies
\begin{align*}
I_{a,b}(x y) = I_{a,b}(x)y + x \, \textnormal{ad}_{-b}(y)
\end{align*}
\noindent
for all $x,y \in {\cal A}$. This fact motivated Bre\v{s}ar to formulate the~following definition.

A linear mapping $u: {\cal A} \rightarrow {\cal A}$ is called a~\textit{generalized derivation} if there exists $d \in \textnormal{Der}({\cal A})$ such that the~\textit{generalized Leibniz rule}
\begin{align}
\label{Leibniz}
u(a b) = u(a) b + a \, d(b) \tag*{(1)}
\end{align}
\noindent
holds for all $a,b \in \cal A$. Derivation $d$ in the~preceding definition is called \textit{associated} with $u$. If such a~derivation is unique, it is written as $d_u$.

The set of all generalized derivations of $\cal A$ is denoted $\textnormal{GDer}({\cal A})$.

The concept of generalized derivation covers the~notion of a~derivation and that of a~\textit{left multiplier}, that is, a~linear map ${\cal L}: {\cal A} \rightarrow {\cal A}$ satisfying ${\cal L}(ab) = {\cal L}(a)b$ for all $a,b \in {\cal A}$. In fact, one can show that any $u \in \textnormal{GDer}({\cal A})$ is a~sum of a~left multiplier and a~derivation associated with $u$. If this decomposition is unique, the~left multiplier ${\cal L}_u = u - d_u$ will also be called \textit{associated} with $u$.

Simple examples of left multipliers include the~maps $l_a$ defined as $l_a(b) = ab$ for any $b \in {\cal A}$. Left multipliers of this form we shall call \textit{inner}. Another important example of a~left multiplier is the~identity map $\textnormal{id}_{\cal A}$. For algebras without left unity, $\textnormal{id}_{\cal A}$ is not inner.

One can easily prove that generalized derivations satisfy (i--iii). However, in the~case of some algebras (iv) does not hold for all generalized derivations. For our later geometrical applications it is important to single out those elements of $\textnormal{GDer}({\cal A})$ for which (iv) holds.

By $\textnormal{CGDer}({\cal A})$ we shall denote the~set of generalized derivations of $\cal A$ that leave ${\cal Z}({\cal A})$ invariant. One can easily check that this set satisfies all properties (i--iv) and is a~proper superset of $\textnormal{Der}({\cal A})$, because $\textnormal{id}_{\cal A} \in \textnormal{CGDer}({\cal A}) \setminus \textnormal{Der}({\cal A})$.

\section{Generalized derivation-based differential geometry}
\label{sec3}

In this section we construct elements of differential geometry based on generalized derivations; we adopt the~method analogous to what is done in similar situations \cite{Sasin,Heller,Parfionov}. The interested reader is referred also to works by Dubois-Violette \cite{Dubois-Violette1,Dubois-Violette2,Djemai}.

For the~sake of readability, let us denote the~${\cal Z}({\cal A})$-module $\textnormal{CGDer}({\cal A})$ simply by $V$. Then $V^\ast \equiv \textnormal{Hom}_{{\cal Z}({\cal A})}\left(V, {\cal Z}({\cal A})\right)$ is its dual ${\cal Z}({\cal A})$-module.

Let ${\cal G}: V \times V \rightarrow {\cal Z}({\cal A})$ be a~symmetric, ${\cal Z}({\cal A})$-bilinear map called \textit{metric}. We will also assume that ${\cal G}$ is nondegenerate, that is, that the~map $\Phi_{{\cal G}}:V\rightarrow V^\ast$ given by
\begin{align*}
\Phi_{{\cal G}}(u)(v)={\cal G}(u,v)
\end{align*}
\noindent
is an~isomorphism of ${\cal Z}({\cal A})$-modules.

We are now ready to define the~\textit{preconnection} $\nabla^\ast: V \times V \rightarrow V^\ast$ by using the~Koszul formula \cite{Heller}
\begin{align*}
\left(\nabla^\ast_u v \right)(w) = & \, \tfrac{1}{2} \left[ u\left({\cal G} (v,w)\right) + v\left({\cal G} (u,w)\right) - w\left({\cal G} (u,v)\right)\right.
\\ & \left. + \, {\cal G}(w,[u,v]) + {\cal G}(v,[w,u]) - {\cal G}(u, [v,w]) \right]
\end{align*}
\noindent
and then the~\textit{Levi-Civita connection} $\nabla: V \times V \rightarrow V$ by
\begin{align*}
\nabla = \Phi_{{\cal G}}^{-1} \circ \nabla^{\ast}.
\end{align*}

As one can show by tedious but straightforward calculations, $\nabla$ has almost identical properties to its well-known derivation-based counterpart, the~only difference lying in the~generalized Leibniz rule, we thus have
\begin{align*}
& 1^\circ \quad \nabla_{u_1 + u_2} v = \nabla_{u_1} v + \nabla_{u_2} v,
\\
& 2^\circ \quad \nabla_{f u} v = f \, \nabla_u v,
\\
& 3^\circ \quad \nabla_u (v_1 + v_2) = \nabla_u v_1 + \nabla_u v_2,
\\
& 4^\circ \quad \nabla_u (f v) = d_u(f) v + f \, \nabla_u v
\\
& \qquad (\textit{generalized Leibniz rule}),
\\
& 5^\circ \quad \nabla_u v - \nabla_v u - [u, v] = 0
\\
& \qquad (\textit{torsion-freeness}),
\\
& 6^\circ \quad w\left({\cal G} (u,v)\right) = {\cal G} \left( \nabla_w u , v \right) + {\cal G} \left( u, \nabla_w v \right)
\\
& \qquad (\textit{metric compatibility})
\end{align*}
\noindent
for all $u, u_1, u_2, v, v_1, v_2, w \in V$ and $f \in {\cal Z}({\cal A})$. Moreover, the~Levi-Civita connection is the~unique connection satisfying $5^\circ$ and $6^\circ$.

Having defined the~Levi-Civita connection, one can readily introduce the~\textit{Riemann curvature map} $R: V \times V \times V \rightarrow V$ by
\begin{align*}
R(u,v)w = \nabla_u \nabla_v w - \nabla_v \nabla_u w - \nabla_{[u,v]}w.
\end{align*}

It is not difficult to check that, although the~property $4^\circ$ differs from the~ordinary Leibniz rule for connections, the~map $R$ is ${\cal Z}({\cal A})$-trilinear and thus it can be called the~\textit{Riemann tensor}.

$R$ can be shown to satisfy the~usual Riemann tensor identities
\begin{align*}
& R(u,v) = -R(v,u),
\\
& {\cal G} \left( R(u,v)w, z \right) = - {\cal G} \left( R(u,v)z, w \right),
\\
& R(u,v)w + R(v,w)u + R(w,u)v = 0,
\\
& {\cal G} \left( R(u,v)w, z \right) = {\cal G} \left( R(w,z)u, v \right)
\end{align*}
\noindent
for all $u, v, w, z \in V$.

By demanding the~${\cal Z}({\cal A})$-module $\textnormal{CGDer}({\cal A})$ to be (at least locally) free, one can define the~Ricci 2-form and the~scalar curvature using standard construction involving the~notion of a~trace of an~operator. With this in mind, let us move to an~illustrative example of a~(commutative) algebra, whose generalized derivations will be shown to possess interesting physical interpretation.

\section{Generalized derivations of the~algebra of smooth functions}
\label{sec4}

Let $M$ be an~$N$-dimensional lorentzian manifold (we assume $N \geq 2$) and let us consider the~algebra ${\cal A} = C^\infty(M)$ of smooth real-valued functions on $M$ with the~pointwise multiplication.

The set of derivations on $\cal A$ is a~locally free $\cal A$-module and
\begin{align*}
\textnormal{Der}({\cal A}) = \textnormal{span}_{\cal A}\left( \partial_0, \partial_1, \ldots, \partial_{N-1} \right),
\end{align*}
\noindent
where $\partial_\mu \equiv \frac{\partial}{\partial x^\mu}$ in a~fixed map $x = (x^0, x^1, \ldots , x^{N-1})$.

Because ${\cal Z}({\cal A}) = {\cal A}$, all generalized derivations trivially leave the~center invariant, $\textnormal{CGDer}({\cal A}) = \textnormal{GDer}({\cal A})$. In order to find the~local basis of $\textnormal{GDer}({\cal A})$, notice that, by the~generalized Leibniz rule \ref{Leibniz},
\begin{align}
\begin{split}
\label{uf1}
u(f) & = u({\bf 1} \cdot f) = u({\bf 1}) f + d_u(f)
\\
& = {\cal L}_u({\bf 1}) f + d_u(x^\mu) \partial_\mu f
\end{split}
\tag*{(2)}
\end{align}
\noindent
for any $u \in \textnormal{GDer}({\cal A})$ and $f \in {\cal A}$, where ${\bf 1}$ denotes a~constant function equal to one. Thus
\begin{align}
\label{GDerspan}
\textnormal{GDer}({\cal A}) = \textnormal{span}_{\cal A}\left( \partial_0, \partial_1, \ldots, \partial_{N-1}, \textnormal{id}_{\cal A} \right). \tag*{(3)}
\end{align}

Therefore, $\textnormal{dim} \, \textnormal{GDer}({\cal A}) = 1 + \textnormal{dim} \, M$.

For the~sake of brevity let us denote $\partial_N \equiv \textnormal{id}_{\cal A}$. In what follows we shall also adopt the~convention that capital Latin indices $A,B,C,\ldots$ run from 0 to $N$, whereas Greek indices $\mu, \nu, \alpha, \beta, \ldots$ do not cover the~additional ``generalized'' index value $N$. Thus, \ref{uf1} can be re-expressed as
\begin{align*}
u(f) = u^A \partial_A f
\end{align*}
\noindent
with $u^\mu = d_u(x^\mu) = u(x^\mu) - u({\bf 1})x^\mu$ and $u^N = {\cal L}_u({\bf 1}) = u({\bf 1})$.

Of course, coordinate transformations affect all index values but $N$.

Setting $g_{A B} \equiv {\cal G} \left( \partial_A, \partial_B \right)$, we use the~Koszul formula to express the~coefficients of the~Levi-Civita connection
\begin{align}
\begin{split}
\label{Christoffel}
&\nabla_{\partial_A}\partial_B = \Gamma^C_{\ A B} \partial_C \ \quad \textnormal{where}
\\
&\Gamma^C_{\ A B} = \tfrac{1}{2} g^{C D}\left( \partial_A g_{B D} + \partial_B g_{A D} - \partial_D g_{A B} \right).
\end{split}
\tag*{(4)}
\end{align}

Although the~preceding expressions are identical to those known from the~pseudo-Riemannian geometry, the~connection acts in a~slightly different way because of the~presence of $\textnormal{id}_{\cal A}$ in the~basis
\begin{align*}
\nabla_{u^A \partial_A} (v^B \partial_B) = \left( u^A v^B \Gamma^C_{\ A B} + u^\mu \partial_\mu v^C \right) \partial_C.
\end{align*}
\noindent
Notice that among the~indices used here one is Greek.

This can be written in the~abstract index notation as follows:
\begin{align}
\begin{split}
\label{connection_action}
\nabla_{A} v^B = \left( \partial_A - \delta_A^N \right) v^B + \Gamma^B_{\ A C} v^C.
\end{split}
\tag*{(5)}
\end{align}

Let us now consider the~coefficients $R^C_{\ D A B}$ of the~Riemann tensor, defined by the~equality
\begin{align*}
R(\partial_A, \partial_B) \partial_D = R^C_{\ D A B} \partial_C.
\end{align*}

Using \ref{connection_action}, one obtains the~following formula for these coefficients:
\begin{align}
\begin{split}
\label{Riemann}
R^C_{\ D A B} = & \left( \partial_A - \delta_A^N \right) \Gamma^C_{\ B D} - \left( \partial_B - \delta_B^N \right) \Gamma^C_{\ A D} \, +
\\
& + \, \Gamma^K_{\ B D} \Gamma^C_{\ A K} - \Gamma^K_{\ A D} \Gamma^C_{\ B K}.
\end{split}
\tag*{(6)}
\end{align}

Note that \ref{Riemann} differs from the~standard result if $A$ or $B$ is equal to $N$.

As for the~coefficients of the~Ricci 2-form $\textbf{ric}$ and the~scalar curvature $r$, we have, as usual,
\begin{align}
\label{Ricci}
\textbf{ric}_{A B} = R^C_{\ A C B} \qquad \textnormal{and} \qquad r = g^{A B} \textbf{ric}_{A B}. \tag*{(7)}
\end{align}

Let us now visualize the~effect the~introduction of generalized derivations has on Christoffel symbols, on Riemann and Ricci tensors' coefficients and on the~scalar curvature, by calculating them for a~simple (but nontrivial) metric.

\section{Example: a~simple metric}
\label{sec5}

In this section, we consider a~metric that does not mix de\-ri\-va\-tions $\partial_\mu$ with the~identity $\partial_N$. To do so, let us take any $N$-dimensional metric $g_{\alpha \beta}$ and let us set
\begin{align}
\label{metryczka}
g_{A B} =
\left[\begin{array}{cccc}
    & & &\multicolumn{1}{|c}{\ } \\ [-9pt]
    \multicolumn{3}{c}{\multirow{3}{*}{$g_{\alpha \beta}$}} &
    \multicolumn{1}{|c}{\ 0} \\ [-2pt]
    & & &\multicolumn{1}{|c}{\ \vdots} \\ [-8pt]
    & & &\multicolumn{1}{|c}{\ }\\ [-5pt]
    & & &\multicolumn{1}{|c}{\ 0} \\ [-12pt]
    & & &\multicolumn{1}{|c}{\ }\\ \cline{1-3}
    \\ [-10pt]
   \ 0 & \ldots & 0 \ &
   \ \varepsilon \Phi^2 \\
  \end{array}\right],
\tag*{(8)}
\end{align}
\noindent
where $\Phi = \Phi(x^0, x^1, \ldots, x^{N-1})$ denotes a~smooth positive function and $\varepsilon = \pm 1$. For clarity, we separate the~parts of matrices associated with the~additional ``generalized'' degree of freedom from the~``classical'' $N \times N$ parts.

In the~following, the~tilde above a~given object signifies that the~object is obtained from the~$N$-dimensional metric $g_{\alpha \beta}$ according to the~standard (i.e. not ``generalized'') pseudo-rie\-man\-nian-geometrical formulae.

One obtains the~following Christoffel symbols:
\begin{align*}
&\Gamma^\gamma_{\ \alpha \beta}  = \tfrac{1}{2}g^{\gamma \lambda}\left( \partial_\alpha g_{\beta \lambda} + \partial_\beta g_{\alpha \lambda} - \partial_\lambda g_{\alpha \beta} \right) = \widetilde{\Gamma}^\gamma_{\ \alpha \beta},
\\
&\Gamma^N_{\ \alpha \beta}  = -\frac{\varepsilon}{2 \Phi^2} g_{\alpha \beta},
\\
&\Gamma^\gamma_{\ N \beta}  = \Gamma^\gamma_{\ \beta N} = \tfrac{1}{2}\delta^\gamma_\beta,
\\
&\Gamma^\gamma_{\ NN}  = -\varepsilon \Phi \partial^\gamma \Phi,
\\
&\Gamma^N_{\ \alpha N}  = \Gamma^N_{\ N \alpha} = \frac{\partial_\alpha \Phi}{\Phi},
\\
&\Gamma^N_{\ NN} = \tfrac{1}{2}
\end{align*}
\noindent
where $\partial^\gamma \equiv g^{\gamma \lambda} \partial_\lambda$.

As for the~nonzero Riemann tensor coefficients, one gets
\begin{align*}
&R^\gamma_{\ \lambda \alpha \beta}  = \widetilde{R}^\gamma_{\ \lambda \alpha \beta} + \frac{\varepsilon}{4 \Phi^2}(g_{\alpha \lambda} \delta^\gamma_\beta - g_{\beta \lambda} \delta^\gamma_\alpha),
\\
&R^N_{\ \lambda \alpha \beta}  = \frac{\varepsilon}{2 \Phi^3} (g_{\beta \lambda}\partial_\alpha - g_{\alpha \lambda}\partial_\beta)\Phi,
\\
&R^\gamma_{\ N \alpha \beta}  = \frac{1}{2 \Phi} (\delta^\gamma_\alpha \partial_\beta - \delta^\gamma_\beta \partial_\alpha)\Phi,
\\
&R^\gamma_{\ \lambda N \beta}  = - R^\gamma_{\ \lambda \beta N} = \frac{1}{2 \Phi} (g_{\beta \lambda} \partial^\gamma - \delta^\gamma_\beta \partial_\lambda)\Phi,
\\
&R^N_{\ \lambda \alpha N}  = - R^N_{\ \lambda N \alpha} = \frac{1}{\Phi} \widetilde{\nabla}_\alpha \partial_\lambda \Phi,
\\
&R^\gamma_{\ NN \beta}  = - R^\gamma_{\ N \beta N} = \varepsilon \Phi \widetilde{\nabla}_\beta \partial^\gamma \Phi
\end{align*}
\noindent
where $\widetilde{\nabla}_\alpha$ denotes the~standard covariant derivative along the~$\alpha$-th direction.

The Ricci tensor, written in matrix form, reads
\begin{align*}
\textbf{ric}_{A B} = \left[\begin{array}{cc}
& \multicolumn{1}{|c}{\ } \\ [-8pt]
\widetilde{\textbf{ric}}_{\alpha \beta} - \varepsilon \frac{N - 1}{4 \Phi^2} g_{\alpha \beta} - \frac{1}{\Phi} \widetilde{\nabla}_\alpha \partial_\beta \Phi &
\multicolumn{1}{|c}{\frac{N - 1}{2 \Phi} \partial_\alpha \Phi}
\\ [8pt]
\cline{1-1}
\\ [-8pt]
\frac{N - 1}{2 \Phi} \partial_\beta \Phi &
-\varepsilon \Phi \widetilde{\Delta} \Phi \\
  \end{array}\right]
\end{align*}
\noindent
where $\widetilde{\Delta}$ denotes the~standard Laplace--Beltrami operator
\begin{align*}
\widetilde{\Delta} = \widetilde{\nabla}_\lambda \partial^\lambda = \partial^\lambda \partial_\lambda - g^{\mu \nu} \widetilde{\Gamma}^\lambda_{\ \mu \nu} \partial_\lambda.
\end{align*}

Finally, the~scalar curvature takes the~following form
\begin{align*}
r = \widetilde{r} - \varepsilon \frac{N(N - 1)}{4 \Phi^2} - \frac{2}{\Phi} \widetilde{\Delta} \Phi.
\end{align*}

Notice that the~introduction of the~generalized derivation $\partial_N$ leads to the~appearance of additional terms depending on $\Phi$, similar to when considering an~extra dimension within Kaluza--Klein theories (see section \ref{sec7} for the~list of references). We shall investigate this similarity more closely in section \ref{sec7}.

\section{Action principle and generalized Einstein's equations}
\label{sec6}

We have now all the~geometric notions needed to investigate the~impact the~generalized derivations have on Einstein's field equations. Let us start with the~following Einstein--Hilbert action:
\begin{align}
\label{action}
S_{EH} = \frac{1}{2 \kappa} \int r \sqrt{|g|} \, d^N x,
\tag*{(9)}
\end{align}
\noindent
where $g$ denotes the~determinant of the~$(N+1) \times (N+1)$ matrix $g_{A B}$. The coefficient $\kappa$ is a~physical constant equal to $\frac{8 \pi G}{c^4}$, where $G$ is the~gravitational constant and $c$ is the~speed of light.

One should notice that the~term $\sqrt{|g|} \, d^N x$ differs from the~standard volume $N$-form, which would involve only the~determinant of the~``classical'' $N \times N$ part of the~matrix $g_{A B}$. Nevertheless, just as the~standard volume $N$-form, the~term $\sqrt{|g|} \, d^N x$ is invariant under coordinate transformations.

Let us vary thus defined $S_{EH}$ with respect to $\delta g^{A B}$
\begin{align}
\begin{split}
\label{variation1}
\delta S_{EH} = & \ \frac{1}{2 \kappa} \int \left( \textbf{ric}_{A B} - \frac{1}{2} r g_{A B} \right) \delta g^{A B} \sqrt{|g|} \, d^Nx
\\
& + \ \frac{1}{2 \kappa} \int \delta \textbf{ric}_{A B} g^{A B} \sqrt{|g|} \, d^Nx.
\end{split}
\tag*{(10)}
\end{align}

The integrand involving $\delta \textbf{ric}_{A B}$ does not vanish and can be expressed via the~variations of Christoffel symbols as follows:
\begin{align}
\begin{split}
\label{variation2}
\delta \textbf{ric}_{A B} g^{A B} \sqrt{|g|}
& \ = \partial_\lambda \left[ \left(g^{A B} \delta \Gamma^\lambda_{\ A B} - g^{\lambda B} \delta \Gamma^{A}_{\ A B}\right) \sqrt{|g|} \right]
\\
& \ \quad + \, \tfrac{N-1}{2} \left(g^{A B} \delta \Gamma^N_{\ A B} - g^{N B} \delta \Gamma^{A}_{\ A B}\right) \sqrt{|g|}.
\end{split}
\tag*{(11)}
\end{align}

The first term on the~right-hand side of \ref{variation2} is a~divergence and as such can be omitted in further considerations.

In order to express the~remaining term with $\delta g^{A B}$, let us notice that
\begin{align*}
&g^{A B} \Gamma^N_{\ A B} - g^{N B} \Gamma^{A}_{\ A B} = -N g^{N N} - \partial_\lambda g^{N \lambda} - g^{N \lambda} g^{A B} \partial_\lambda g_{A B}.
\end{align*}

By varying the~preceding equality, one gets
\begin{align}
\begin{split}
\label{variation3}
g^{A B} \delta \Gamma^N_{\ A B} - g^{N B} \delta \Gamma^{A}_{\ A B}
& \ = \Gamma^{A}_{\ A B} \delta g^{N B} - \Gamma^N_{\ A B} \delta g^{A B} - N \delta g^{N N}
\\
& \ \quad - \partial_\lambda \delta g^{N \lambda} - g^{A B} \partial_\lambda g_{A B} \delta g^{N \lambda}
\\
& \ \quad - g^{N \lambda} \partial_\lambda g_{A B} \delta g^{A B} - g^{N \lambda} g^{A B} \partial_\lambda \delta g_{A B}.
\end{split}
\tag*{(12)}
\end{align}

Inserting this expression into \ref{variation2}, one can further simplify it by first realizing that
\begin{align*}
\Gamma^{A}_{\ A B} \delta g^{N B} = \frac{1}{2} g^{AC} \partial_B g_{AC} \delta g^{NB} = \frac{1}{2} g^{AC} \partial_{\lambda} g_{AC} \delta g^{N \lambda} + \frac{N+1}{2} \delta g^{NN}
\end{align*}
\noindent
and that
\begin{align*}
-\partial_\lambda \delta g^{N \lambda} \sqrt{|g|} = -\partial_\lambda \left( \delta g^{N \lambda} \sqrt{|g|} \right) + \frac{1}{2} g^{CD} \partial_\lambda g_{CD} \sqrt{|g|} \, \delta g^{N \lambda}.
\end{align*}

Therefore, up to divergence terms
\begin{align}
\begin{split}
\label{variation4}
&\big( \Gamma^{A}_{\ A B} \delta g^{N B} - N \delta g^{N N} - \partial_\lambda \delta g^{N \lambda} - g^{A B} \partial_\lambda g_{A B} \delta g^{N \lambda} \big) \sqrt{|g|}
\\
& \ = - \frac{N-1}{2} \sqrt{|g|} \, \delta g^{N N}.
\end{split}
\tag*{(13)}
\end{align}

Let us now move to the~two rightmost terms in \ref{variation3}. Because it is true
that\footnote{To prove this claim, one can use the~formula for the~derivative of the~matrix inverse, obtaining
$$\delta g_{AB} \partial_{\lambda} g^{AB} = \left(- g_{AC} \delta g^{CD} g_{DB} \right) \partial_{\lambda} g^{AB} = \delta g^{CD} \left(- g_{CA} \partial_{\lambda} g^{AB} g_{BD} \right) = \delta g^{CD} \partial_{\lambda} g_{CD}.$$}
\begin{align*}
\delta g_{AB} \partial_{\lambda} g^{AB} = \delta g^{AB} \partial_{\lambda} g_{AB},
\end{align*}
\noindent
one can write that
\begin{align*}
&- g^{N \lambda} \partial_\lambda g_{A B} \delta g^{A B} - g^{N \lambda} g^{A B} \partial_\lambda \delta g_{A B}
\\
&\ = - g^{N \lambda} \partial_\lambda g^{A B} \delta g_{A B} - g^{N \lambda} g^{A B} \partial_\lambda \delta g_{A B}
\\
&\ = - g^{N \lambda} \partial_\lambda \left( g^{A B} \delta g_{A B} \right) = g^{N \lambda} \partial_\lambda \left( g_{A B} \delta g^{A B} \right).
\end{align*}

This means, however, that
\begin{align}
\begin{split}
\label{variation5}
& \big( - g^{N \lambda} \partial_\lambda g_{A B} \delta g^{A B} - g^{N \lambda} g^{A B} \partial_\lambda \delta g_{A B} \big) \sqrt{|g|}
\\
& \ = g^{N \lambda} \sqrt{|g|} \, \partial_\lambda \left( g_{A B} \delta g^{A B} \right)
\\
& \ = - \partial_\lambda \left( g^{N \lambda} \sqrt{|g|} \right) g_{A B} \delta g^{A B}
\end{split}
\tag*{(14)}
\end{align}
\noindent
up to divergence terms.

All in all, \ref{variation2}--\ref{variation5} imply that up to divergence terms
\begin{align}
\begin{split}
\label{variation6}
\delta \textbf{ric}_{A B} g^{A B} \sqrt{|g|}
& \ = \frac{N-1}{2} \bigg( - \Gamma^N_{\ A B} \sqrt{|g|} \, \delta g^{A B} - \frac{N-1}{2} \sqrt{|g|} \, \delta g^{N N} \bigg.
\\
& \hspace{2.15cm} \bigg. - \partial_\lambda \left( g^{N \lambda} \sqrt{|g|} \right) g_{A B} \delta g^{A B}\bigg).
\end{split}
\tag*{(15)}
\end{align}

Therefore, the~variation of \ref{variation1} can finally be put into the~following form:
\begin{align}
\begin{split}
\label{variation7}
\delta S_{EH} = \frac{1}{2 \kappa} \int & \left( \textbf{ric}_{A B} - \frac{1}{2} r g_{A B} - \frac{N-1}{2} \Pi_{A B} \right) \delta g^{A B} \sqrt{|g|} \, d^Nx
\end{split}
\tag*{(16)}
\end{align}
\noindent
where
\begin{align}
\begin{split}
\label{variation8}
\Pi_{A B} & = \Gamma^N_{\ A B} + \frac{N-1}{2} \delta_A^N \delta_B^N + g_{A B} \frac{1}{\sqrt{|g|}} \partial_\lambda \left( g^{N \lambda} \sqrt{|g|} \right)
\\
& = \Gamma^N_{\ A B} + \frac{N-1}{2} \delta_A^N \delta_B^N + g_{A B} \sqrt{|g^{NN}|} \, \widetilde{\nabla}_\lambda \left( \frac{g^{N \lambda}}{\sqrt{|g^{NN}|}} \right)
\end{split}
\tag*{(17)}
\end{align}
\noindent
is a~symmetric tensor. Notice that it naturally involves the~term proportional to the~metric\footnote{Also, an~additional term of this kind is possibly implicit in $\Gamma^N_{\ A B}$, as for the~case of the~metric discussed in section \ref{sec5}.}. It is thus tempting to associate it with Einstein's cosmological term $\Lambda  g_{\alpha \beta}$ with nonconstant $\Lambda$ which could model dynamical dark energy, here being of a~purely geometrical (or rather generalized-geometrical) origin.

Let now $S_M$ denote the~standard action for matter. Applying the~action principle to the~sum $S_{EH} + S_M$ one gets the~following generalized Einstein's equations
\begin{align}
\label{Einstein1}
& \textbf{ric}_{A B} - \frac{1}{2} r g_{A B} - \frac{N-1}{2} \Pi_{A B} = \kappa \sqrt{|g^{NN}|} \, T_{A B},
\tag*{(18)}
\end{align}
\noindent
where the~stress--energy tensor with indices raised $T^{\alpha \beta}$ is given as usual by
\begin{align*}
T^{A B} = - \frac{2}{\sqrt{-\widetilde{g}}} \frac{\delta S_M}{\delta g_{A B}}.
\end{align*}

Because we have assumed that $S_M$ is standard, that is, it does not involve metric coefficients $g_{\alpha N}, g_{N \beta}, g_{NN}$, the~stress--energy tensor will be of the~form
\begin{align*}
T^{A B} =
\left[\begin{array}{cccc}
    & & &\multicolumn{1}{|c}{\ } \\ [-9pt]
    \multicolumn{3}{c}{\multirow{3}{*}{$T^{\alpha \beta}$}} &
    \multicolumn{1}{|c}{\ 0} \\ [-2pt]
    & & &\multicolumn{1}{|c}{\ \vdots} \\ [-8pt]
    & & &\multicolumn{1}{|c}{\ }\\ [-5pt]
    & & &\multicolumn{1}{|c}{\ 0} \\ [-12pt]
    & & &\multicolumn{1}{|c}{\ }\\ \cline{1-3}
    \\ [-10pt]
   \ 0 & \ldots & 0 \ &
   \ 0 \\
  \end{array}\right]
\end{align*}
\noindent
and so $T_{A B} = g_{A \alpha} g_{B \beta} T^{\alpha \beta}$. Notice the~equality of the~traces
\begin{align*}
T^A_{\ A} = g^{A B} g_{A \alpha} g_{B \beta} T^{\alpha \beta} = g_{\alpha \beta} T^{\alpha \beta} = T^\alpha_{\ \alpha}.
\end{align*}

By calculating the~trace of both sides of \ref{Einstein1} one obtains
\begin{align}
\label{Einstein2}
r + \frac{N}{\sqrt{|g|}} \partial_\alpha \left( \sqrt{|g|} g^{N \alpha} \right) = -\frac{2 \kappa}{N - 1} \sqrt{|g^{N N}|} \, T^\alpha_{\ \alpha}.
\tag*{(19)}
\end{align}

It is appropriate to call the~left-hand side of \ref{Einstein1} the~\textit{generalized Einstein tensor}. What is worth noticing is that it involves no free parameters.

The term $\sqrt{|g^{NN}|}$ on the~right-hand sides of \ref{Einstein1} and \ref{Einstein2} is in general nonconstant; it models the~space--time-dependency of the~gravitational constant $G$, similarly to the~Brans--Dicke scalar field \cite{Brans}.

In fact, by considering only the~metrics studied in section \ref{sec5}, that is, those for which $g_{\alpha N} = g_{N \alpha} = 0$ and $g_{NN} = \varepsilon \Phi^2$ (where $\varepsilon = \pm 1$), the~theory reduces exactly to the~scalar--tensor theory governed by the~action
\begin{align}
\label{OH}
S_{O'H} = \frac{1}{2 \kappa} \int \left( \Phi \widetilde{r} - V[\Phi] \right) \sqrt{-\widetilde{g}} \, d^Nx
\tag*{(20)}
\end{align}
with the~``Coulomb'' potential $V[\Phi] = \varepsilon \frac{N(N-1)}{4 \Phi}$.

An action of this kind was first considered by O'Hanlon \cite{OHanlon} and is sometimes referred to as the~O'Hanlon action (consult Sotiriou and Faraoni \cite{Sotiriou} for more references). It is a~special case of the~Brans--Dicke action with a~nonzero potential and the~Brans--Dicke parameter $\omega$ equal to zero (in the~Jordan frame).

One can also put \ref{OH} into an~equivalent, $f(R)$-the\-o\-ret\-i\-cal form\footnote{Actions considered in $f(R)$-theory of gravity have the~general form $$S = \frac{1}{2 \kappa} \int f\left(\widetilde{r}\right) \sqrt{-\widetilde{g}} \, d^Nx.$$} \cite{Sotiriou, Faraoni1}
\begin{align*}
S_{f(R)} = \frac{\sqrt{N(N-1)}}{2 \kappa} \int \sqrt{ |\widetilde{r}| } \, \sqrt{-\widetilde{g}} \, d^Nx.
\end{align*}

Thus, from the~point of view of $f(R)$-gravity theory, the~introduction of generalized derivations (and considering only the~narrowed class of metrics) leads to the~action with $f(R) = \sqrt{N(N-1)} |R|^{1/2}$. Actions of the~form $f(R) \propto |R|^n$ (with $n$ not necessarily integer) have been studied by numerous authors; see Faraoni \cite{Faraoni2} for a~list of references. It is already known that only for $n \approx 1$ with the~level of accuracy of about $10^{-19}$ theories of this type meet current observational data as far as Solar System physics is concerned \cite{Faraoni2}.

\section{Noncompact invisible dimension}
\label{sec7}

The idea of the~celebrated Kaluza--Klein theory (and its modifications) is to assume that the~physical space--time is a~pseudo-Riemannian manifold of dimension $D>4$, on which one considers the~Einstein--Hilbert action
\begin{align}
\label{KKaction}
S_{KK} = \frac{1}{2 \kappa} \int \widehat{r} \sqrt{|\widehat{g}|} \, d^D x
\tag*{(21)}
\end{align}
\noindent
where $\widehat{g}$ denotes the~determinant of the~$D \times D$ matrix of the~metric tensor $\widehat{g}_{A B}$ and $\widehat{r}$ is the~scalar curvature obtained from that metric with the~standard pseudo-Riemannian-geometric formulae.

Kaluza \cite{Kaluza} showed that in $D=5$ dimensions varying \ref{KKaction} over the~class of metrics satisfying the~so-called ``cylinder condition'' (i.e. the~class of metrics independent of the~extra coordinate $x_4$) leads to a~theory unifying Einstein's theory of gravity with Maxwell's theory of electromagnetism. Shortly after, Klein \cite{Klein} realized that if the~extra dimension is compact and has a~small enough scale, the~seemingly artificial ``cylinder condition'' arises naturally in the~low-energy physics regime. Moreover, the~previously mentioned features of the~fifth dimension explain why it is not physically observed \cite{Bergmann}.

Throughout the~decades, numerous authors introduced and studied various modifications and generalizations of Kaluza's original idea, in order to incorporate other physical phenomena into a~unified geometrical formalism. Because it is far beyond the~scope of this work even to briefly describe them, the~interested reader is referred to excellent books \cite{Book1,Book2,Book3,Book4,Book5} and review papers \cite{Wesson,Toms,Appelquist,Duff1,Duff2,Bailin}.
\\

It turns out that the~generalized derivation-based approach to general relativity presented in previous sections can be equivalently formulated in a~Kaluza-Klein-theoretical way.

Concretely, we shall prove that generalized Einstein's equations \ref{Einstein1} with $T_{A B} = 0$ can be obtained from varying Kaluza--Klein action \ref{KKaction} involving one \textit{noncompact} extra dimension ($D = N+1$) over the~class of metrics satisfying the~``modified cylinder condition''. Namely, the~metrics are assumed to depend on the~extra coordinate $x_N$ exponentially, that is
\begin{align}
\label{cylinder}
\widehat{g}_{A B} = e^{x_N} g_{A B}
\tag*{(22)}
\end{align}
\noindent
where $g_{A B}$ already \textit{does not} depend on $x_N$. According to the~authors' best knowledge, this particular version of Kaluza--Klein theory has not so far been studied.

To start the~proof, notice that the~determinants of these two matrices satisfy the~equality
\begin{align*}
\widehat{g} = e^{(N+1)x_N} g,
\end{align*}
\noindent
which can be inserted into \ref{KKaction}. We would like to do something similar with $\widehat{r}$. It is crucial to realize that
\begin{align}
\label{claim}
\widehat{r} = e^{-x_N} r
\tag*{(23)}
\end{align}
\noindent
where $r$ is a~scalar curvature obtained from $g_{A B}$ with the~help of formulae \ref{Christoffel}--\ref{Ricci}.

Before we prove this claim, let us introduce the~symbol $\widehat{\partial}_C$ to denote $\frac{\partial}{\partial x^C}$. We use the~hat here so as to avoid confusion, because throughout the~paper the~symbol $\partial_N$ denotes the~identity map $\textnormal{id}_{\cal A}$. Let us also recall that we follow the~convention that capital Latin indices run from 0 to $N$, whereas Greek indices run from 0 to $N-1$.

In order to prove \ref{claim}, let us first notice that
\begin{align}
\label{claim1}
\widehat{\partial}_C \widehat{g}_{A B} = e^{x_N} \partial_C g_{A B}.
\tag*{(24)}
\end{align}

Indeed, by a~simple computation
\begin{align*}
\widehat{\partial}_\lambda \widehat{g}_{A B} = \widehat{\partial}_\lambda \left( e^{x_N} g_{A B} \right) = e^{x_N} \widehat{\partial}_\lambda g_{A B} = e^{x_N} \partial_\lambda g_{A B},
\end{align*}
\noindent
where the~hat can be dropped because for $\lambda = 0,1,\ldots,N-1$ the~meanings of the~symbols $\widehat{\partial}_\lambda$ and $\partial_\lambda$ coincide.

One also obtains
\begin{align*}
\widehat{\partial}_N \widehat{g}_{A B} & \ = \widehat{\partial}_N \left( e^{x_N} g_{A B} \right)
= \widehat{\partial}_N \left( e^{x_N} \right) g_{A B} + e^{x_N} \widehat{\partial}_N \left( g_{A B} \right)
\\
& \ = e^{x_N} g_{A B} = e^{x_N} \partial_N g_{A B},
\end{align*}
\noindent
which proves \ref{claim1}.

The next step is to show that the~coefficients $\widehat{\Gamma}^C_{\ A B}$ of the~Levi-Civita connection associated with $\widehat{g}_{A B}$ do not depend on $x_N$ and are in fact equal to $\Gamma^C_{\ A B}$ given by \ref{Christoffel}. Indeed, by \ref{claim1} one has
\begin{align*}
\widehat{\Gamma}^C_{\ A B} & \ = \tfrac{1}{2} \widehat{g}^{C D}\left( \widehat{\partial}_A \widehat{g}_{B D} + \widehat{\partial}_B \widehat{g}_{A D} - \widehat{\partial}_D \widehat{g}_{A B} \right)
\\
& \ = \tfrac{1}{2} e^{-x_N} g^{C D}\big( e^{x_N} \partial_A g_{B D} + e^{x_N} \partial_B g_{A D} - e^{x_N} \partial_D g_{A B} \big)
\\
& \ = \tfrac{1}{2} g^{C D}\big( \partial_A g_{B D} + \partial_B g_{A D} - \partial_D g_{A B} \big) = \Gamma^C_{\ A B}.
\end{align*}

Since $\widehat{\Gamma}^C_{\ A B}$ does not depend on $x_N$, the~standard formula for the~Riemann tensor coefficients leads to a~formula identical to \ref{Riemann}. Consequently, the~same concerns the~Ricci tensor coefficients, therefore,
\begin{align*}
\widehat{R}^C_{\ D A B} = R^C_{\ D A B} \qquad \textrm{and} \qquad \widehat{\textbf{ric}}_{A B} = \textbf{ric}_{A B}
\end{align*}
\noindent
with the~right-hand sides obtained from $g_{A B}$ via formulae \ref{Riemann} and \ref{Ricci}.

Finally, for the~scalar curvature, one has
\begin{align*}
\widehat{r} = \widehat{g}^{A B} \widehat{\textbf{ric}}_{A B} = e^{-x_N} g^{A B} \textbf{ric}_{A B} = e^{-x_N} r,
\end{align*}
which proves claim \ref{claim}.

With all of this in mind, let us now vary action \ref{KKaction} with respect to $\delta \widehat{g}_{A B} = e^{x_N} \delta g_{A B}$. One obtains
\begin{align}
\nonumber
\delta S_{KK} = & \ \frac{1}{2 \kappa} \int \left( \widehat{\textbf{ric}}_{A B} - \frac{1}{2} \widehat{r} \, \widehat{g}_{A B} \right) \delta \widehat{g}^{A B} \sqrt{|\widehat{g}|} \, d^{N+1}x
\\
\label{KKaction1}
& + \ \frac{1}{2 \kappa} \int \delta \widehat{\textbf{ric}}_{A B} \widehat{g}^{A B} \sqrt{|\widehat{g}|} \, d^{N+1}x.
\tag*{(25)}
\end{align}

The integrand of the~leftmost integral can be equivalently written as
\begin{align*}
& \left( \widehat{\textbf{ric}}_{A B} - \frac{1}{2} \widehat{r} \, \widehat{g}_{A B} \right) \delta \widehat{g}^{A B} \sqrt{|\widehat{g}|} = e^{\tfrac{N-1}{2} x_N} \left( \textbf{ric}_{A B} - \frac{1}{2} r \, g_{A B} \right) \delta g^{A B} \sqrt{|g|}.
\end{align*}

As for the~integrand of the~rightmost integral, it can be expressed as a~$(N+1)$-dimensional divergence
\begin{align*}
\delta \widehat{\textbf{ric}}_{A B} \widehat{g}^{A B} \sqrt{|\widehat{g}|} = \widehat{\partial}_C \left[ \sqrt{|\widehat{g}|} \left( \delta \widehat{\Gamma}^C_{\ AB} \widehat{g}^{AB} - \delta \widehat{\Gamma}^D_{\ DA} \widehat{g}^{AC} \right) \right].
\end{align*}

This, however, \textit{does not} imply that the~rightmost integral in \ref{KKaction1} vanishes. Indeed, because the~Christoffel symbols do not depend on $x_N$, one cannot argue that their variations are zero on a~boundary of a~sufficiently large $(N+1)$-dimensional domain. However, writing down the~dependence on $x_N$ explicitly, one obtains
\begin{align}
\begin{split}
\label{KKaction2}
\delta \widehat{\textbf{ric}}_{A B} \widehat{g}^{A B} \sqrt{|\widehat{g}|}
& \ = \widehat{\partial}_C \left[ e^{\tfrac{N-1}{2} x_N} \sqrt{|g|} \left( \delta \Gamma^C_{\ AB} g^{AB} - \delta \Gamma^D_{\ DA} g^{AC} \right) \right]
\\
& \ = e^{\tfrac{N-1}{2} x_N} \partial_\lambda \left[ \sqrt{|g|} \left( \delta \Gamma^\lambda_{\ AB} g^{AB} - \delta \Gamma^D_{\ DA} g^{A\lambda} \right) \right]
\\
& \ \quad + \frac{N-1}{2} e^{\tfrac{N-1}{2} x_N} \sqrt{|g|} \left( \delta \Gamma^N_{\ AB} g^{AB} - \delta \Gamma^D_{\ DA} g^{AN} \right).
\end{split}
\tag*{(26)}
\end{align}

In fact, this is an~alternate derivation of \ref{variation2}.

By Fubini's theorem and by the~fact that the~variations $\delta \Gamma^C_{\ AB}$ vanish on a~boundary of a~sufficiently large $N$-dimensional subset of any surface of fixed $x_N$, one has
\begin{align*}
\int e^{\tfrac{N-1}{2} x_N} \partial_\lambda \left[ \sqrt{|g|} \left( \delta \Gamma^\lambda_{\ AB} g^{AB} - \delta \Gamma^D_{\ DA} g^{A\lambda} \right) \right] \, d^{N+1} x = 0.
\end{align*}

In other words, one can omit the~first term on the~right-hand side of \ref{KKaction2} and from now on proceed exactly as shown in the~previous section, eventually obtaining Einstein's equations \ref{Einstein1} with zero stress--energy tensor.
\\

This result has an~interesting interpretative aspect. Every Kaluza--Klein model involving noncompact extra dimensions has to address the~question of why these additional dimensions are not observed. For instance Schmutzer's five-dimensional Projective Unified Field Theory (PUFT) \cite{Schmutzer1,Schmutzer2,Schmutzer3} assumes the~additional dimension merely as a~mathematical tool without direct physical meaning. On the~other hand, the~5-dimensional Space--Time--Matter (STM) theory (the interested reader is referred to Overduin and Wesson \cite{Wesson} for a~brief introduction and a~list of references, and to Wesson \cite{WessonBook} for a~more detailed course) treats the~fifth coordinate as physical, but not lengthlike.

Our case in principle seems to resemble Schmutzer's in the~sense that the~extra dimension is nonphysical and effectively results from modifying the~standard pseudo-Riemannian geometry. Indeed, in terms of generalized differential geometry, space--time has an~extra ``generalized-differential dimension'' spanned by $\textrm{id}_{\cal A}$ (see \ref{GDerspan}), which is not associated with any extra space--time coordinate.
\\

It is worth noting that Einstein's equations differ here from the~Ricci-flatness condition $\textbf{ric}_{A B} = 0$ as obtained in other Kaluza--Klein theories without higher-dimensional matter \cite{Wesson}. Indeed, setting $T_{A B} = 0$ in \ref{Einstein1}--\ref{Einstein2} yields
\begin{align}
\label{Einstein3}
& \textbf{ric}_{A B} - \frac{1}{2N} r g_{A B} - \frac{N-1}{2} \left( \Gamma^N_{\ A B} + \frac{N-1}{2} \delta_A^N \delta_B^N \right) = 0,
\tag*{(27)}
\end{align}
\noindent
which does not in general imply that $\textbf{ric}_{A B} = 0$, as one can check for example for metrics discussed in section \ref{sec5}.

One can regard this effect as a~new realization of an~induced matter (``matter-out-of-geometry'') mechanism. In Kaluza's original work, (as well as in the~STM theory \cite{Wesson, WessonBook, Liu}), a~four-dimensional stress--energy tensor appears when the~five-di\-men\-sional Ricci-flatness condition is projected onto the~four-di\-men\-sional setting. In our case additional terms are present in Einstein's equations \ref{Einstein3} already before the~projection. Therefore, we could say that the~``modified cylinder condition'' \ref{cylinder} induces a~certain form of five-dimensional matter.
\\

Let us finally remark, that the~approach based on generalized derivations does not exclude other Kaluza--Klein-type approaches. In other words, \textit{in addition to} treating space--time as a~$D$-dimensional manifold with $D>4$, one can consider its generalized geometry, and interpret it physically if there is such a~necessity.

\end{document}